# Polymer nanofibers as novel light-emitting sources and lasing material

A. Camposeo[a,b], L. Persano[a,b], D. Pisignano*[a,b,c]

[a] National Nanotechnology Laboratory of Istituto Nanoscienze-CNR, via Arnesano, I-73100 Lecce (Italy); [b] Center for Biomolecular Nanotechnologies @UNILE, Istituto Italiano di Tecnologia, via Barsanti, I-73010 Arnesano, LE (Italy); [c] Dipartimento di Matematica e Fisica "Ennio De Giorgi", Università del Salento, via Arnesano I-73100 Lecce (Italy)

## ABSTRACT

Polymer micro- and nano-fibers, made of organic light-emitting materials with optical gain, show interesting lasing properties. Fibers with diameters from few tens of nm to few microns can be fabricated by electrospinning, a method based on electrostatic fields applied to a polymer solution. The morphology and emission properties of these fibers, composed of optically inert polymers embedding laser dyes, are characterized by scanning electron and fluorescence microscopy, and lasing is observed under optical pumping for fluences of the order of $10^2$ μJ cm$^{-2}$. In addition, light-emitting fibers can be electrospun by conjugated polymers, their blends, and other active organics, and can be exploited in a range of photonic and electronic devices. In particular, waveguiding of light is observed and characterized, showing optical loss coefficient in the range of $10^2$-$10^3$ cm$^{-1}$. The reduced size of these novel laser systems, combined with the possibility of achieving wavelength tunability through transistor or other electrode-based architectures embedding non-linear molecular layers, and with their peculiar mechanical robustness, open interesting perspectives for realizing miniaturized laser sources to integrate on-chip optical sensors and photonic circuits.

## 1. INTRODUCTION

In the last decade a lot of experimental efforts have been directed to miniaturize nanoelectronic and optoelectronic components, light-emitting sources, and ultimately laser devices, due the increasing demand of compact photon sources for integrated photon circuits and for lab-on-a-chip devices relying on microfluidic operation. In particular, organic lasers have recently attracted renewed interest as potentially disposable, low-cost, and compact photon sources [1]. Furthermore, evidences of exotic photonic effects in nanostructured systems and organic lasers [2], including the possibility of achieving sub-wavelength optical confinement and random lasing, are further reasons which are motivating research in the field. Organic lasers still need to be powered by external optical excitation. Recent works have demonstrated nanostructured polymer film lasers pumped by UV light emitting diodes and near-infrared solid state laser sources [3,4]. These studies have opened the way to the realization of hybrid laser systems based on well-established, inorganic photon sources as pump, and on flexible organic structures as active material. This approach would allow to exploit the largely tunable optical gain of organics, thus overcoming some limitations of inorganic laser sources that are available only at discrete wavelengths. Organic lasing films are mostly based either on an optically inert polymer doped with laser-dyes or on conjugated polymers exhibiting optical gain, and can be embedded in various optical microcavities, or nanostructured by soft and nanoimprint lithography to realize distributed feedback structures.

Recently, alternative organic lasing materials and systems have emerged, that exploit the confinement of light in micro- and sub-microstructures [5,6]. Indeed, organic light-emitting nanomaterials can constitute an interesting alternative to conventional thin film systems, opening new routes for pushing miniaturization and for achieving new photonic functionalities. These nanomaterials can be realized by different methods including various types of vapor deposition and growth or solution-based approaches, by virtue of the unique ease of processing of polymers and organics. Among other methods, electrospinning is a technique which allows continuous polymer fibers with sub-μm diameter to be fabricated, by means of electrostatic fields used to extrude a polymer solution through a metallic needle [7-9]. The fibers formed upon solvent evaporation during electrospinning are collected on metallic plates in the form of nonwoven mats, or on parallel conductive stripes or rotating collectors as arrays of uniaxially aligned fibers.

*dario.pisignano@unisalento.it; phone +39 0832298104; fax +39 0832298146; www.nanojets.eu

In electrospinning, the first stage of the nanofabrication process generally consists in the formation of a droplet of polymer solution at the spinneret. Following the application of the electric bias, an excess charge is accumulated on the surface of the solution droplet (whose tip takes the shape of the so-called Taylor cone), which finally leads to overcoming the liquid surface tension and to forming a jet of solution in the free space. The concomitant solvent evaporation and dramatic stretching occurring along the path of the jet allow solid nanofibers to be finally produced and collected. The prerequisite for a successful spinning process is given by the viscoelastic properties of the spun solutions which have to show sufficient molecular entanglements. For this reason, producing light-emitting nanofibers by electrospinning is often difficult, due to the poor viscoelastic properties of most of these materials and solutions, and requires a careful choice of both the processing and the solution parameters [10]. A possible option is exploiting the much more favorable viscoelastic properties of solutions made of optically inert polymers such as poly(methylmetacrylate) (PMMA) or polystyrene, and incorporating active organic chromophores in these compounds by preparing blends. Other active materials including light-emitting conjugated polymers can be electrospun as well following a proper process optimization [11]. The technique is highly versatile, allowing fibers with different shapes and composition to be realized, and its throughput is high compared to other methods for nanostructures production. Continuous production runs of many hours are possible for electrospinning at laboratory scale, which makes the process interesting at pre-industrial and possibly industrial scale [12].

Here we report on our work concerning the realization of electrospun active fibers and polymer fibers embedding laser dyes. The fibers have typical diameters of the order of few hundreds of nm. Under pulsed optical pumping, lasing is observed at 580 nm, with typical thresholds in the range 50-150 µJ cm$^{-2}$. Bright light-emission and waveguiding from individual fibers are observed and characterized. In addition, the possibilities opened by these materials are not limited at the microscale or at the single-fiber level. Robust free-standing arrays of nanofibers can be also obtained, in which adjacent nanostructures slightly overlap forming mutual joints which increase the mechanical robustness of the overall produced material. The active nanofibers can be realized by conjugated polymers as well, and used in field-effect transistors. In perspective, electrode-based architectures would allow one to achieve wavelength tunability upon embedding non-linear molecular layers in the devices, as already demonstrated by thin film active media.

## 2. METHODS

In experiments with organic dye compounds, PMMA is used as matrix, whereas different laser dyes are embedded as the active component. Polymer fibers are produced by a home-made electrospinning system comprising a syringe pump, a metallic needle and a voltage supply. Typically, we use a syringe, a spinneret (a 18-27 gauge stainless steel needle), a high voltage power supply providing bias values in the range of 15-30 kV, and an Al foil used as collector (10×10 cm$^2$). This foil can be mounted on an isolating stand positioned at a variable distance (5-30 cm) from the metallic needle constituting the termination of the syringe. The active molecules are dissolved at concentrations in the range of 0.2-0.5% wt in 1-3×10$^{-3}$ M chloroform solutions of PMMA. The solution concentrations and flow rates strongly influence the resulting fibers diameter. In particular, the average fiber diameter decreases by 20%, upon decreasing the solution flow rate from 9 to 0.9 µL min$^{-1}$. Similar nanofabrication experiments are carried out in our laboratory by active polymers suitable for photonics (light-emitting conjugated polymers including the poly[(9,9-dioctylfluorenyl-2,7-diyl)-*co*-(*N,N'*-diphenyl)-*N,N'*di(*p*-butyl-oxy-phenyl)-1,4-diaminobenzene)], the poly[2-methoxy-5-(2-ethylhexyl-oxy)-1,4-phenylene-vinylene] and many others) or for nanoelectronic applications, such as the poly[(vinylidenefluoride-*co*-trifluoroethylene] which is piezoelectric. In addition to random mats deposited on metal foils, nanofibers can be arranged in highly aligned arrays by performing depositions onto a cylindrical collector (diameter = 8 cm) placed at a distance in the range from a few cm to 20 cm from the spinneret. All the electrospinning experiments in this work are performed at room temperature, with about 50% air humidity, collecting non-woven mats of solid nanofibers on various kinds of substrates including quartz for subsequent optical characterization. For each deposition of fibers, we typically prepare reference films spin cast from the same polymer solutions.

Following fabrication, the morphology of fibers is investigated by scanning electron microscopy (SEM) and fluorescence microscopy. The SEM analysis is performed by exploiting either a Raith 150 electron beam system or a FEI Nova NanoSEM 450, operating with an acceleration voltage in the range 5-10 kV. Absorption spectra of the nanofibers are collected by a spectrophotometer (Lambda 950, Perkin Elmer). For the spectroscopic investigation of their lasing emission, fibers deposited on quartz are pumped by either the second (*λ*=532 nm) or the third (*λ*=355 nm) harmonic of a neodymium doped yttrium aluminum garnet laser. The excitation fluences are attenuated by neutral density filters and measured by a calibrated energy meter. The fiber emission is collected by an optical fiber, coupled to a monochromator equipped with a charge coupled device (CCD).

Waveguiding measurements on individual active nanofibers are carried out by continuous-wave micro-photoluminescence (µ-PL), using an inverted microscope (IX71, Olympus) equipped with a 60× oil immersion (numerical aperture =1.42) objective, and a CCD camera. The excitation laser beam is coupled into the microscope objective through a dichroic mirror and then focused on the probed nanofibers by the objective. In this way, part of the light emitted by the active polymer material or embedded chromophores, excited by the tightly focused laser spot, is coupled into the nanofiber and waveguided. The fiber optical losses coefficient is determining upon collecting images of the intensity of emission guided and then scattered into free space by the fiber surface, or at the fiber tip, and analyzing the spatial decay of such signal as a function of the distance of the detected point from the exciting laser spot. The µ-PL is measured upon exciting fibers by a diode laser ($\lambda$=405 nm) and collecting the emission spectra by an optical fiber-coupled monochromator (USB 4000, Ocean Optics).

## 3. RESULTS AND DISCUSSION

Figure 1a shows a typical SEM micrograph of a mat of nanofibers based on PMMA, used as viscoelastic matrix to embedded optically active molecules. These fibers have diameters ranging from few hundreds of nm to few microns, with average values around 0.2-0.5 µm, depending on the used set of electrospinning parameters [13]. Fluorescence microscopy images (Figures 1b-c) clearly show the bright and uniform emission along the fibers. Moreover, the emitted light is effectively confined and waveguided inside the nanostructures, as evidenced by the bright tips of the fibers (Figures 1b-d). As evident by the different colors seen at the fiber body and at the fiber tip in the collected micrographs, the light exiting the tip is always red-shifted with respect to the pristine emission in the nanostructure. This suggests a gradual removal of the high-energy spectral component of light as photons travel along the fiber, which is consistent with residual self-absorption contributing to optical losses. Scattering from the fiber surface or from possible inhomogeneities or aggregates along the fiber is another possible source of optical losses. Typically, however, waveguiding is appreciable also in bent fibers, and cross-talking of adjacent polymer filaments is observed as well, with light emitted in a fiber partially coupling in other fibers nearby, which are not excited directly (Figure 1d). These studies allow optical losses along the longitudinal fiber axis to be measured, and the loss coefficient, $\alpha$, of the waveguides to be estimated from the well-known Beer-Lambert expression:

$$I = I_0 \exp(-\alpha d), \qquad (1)$$

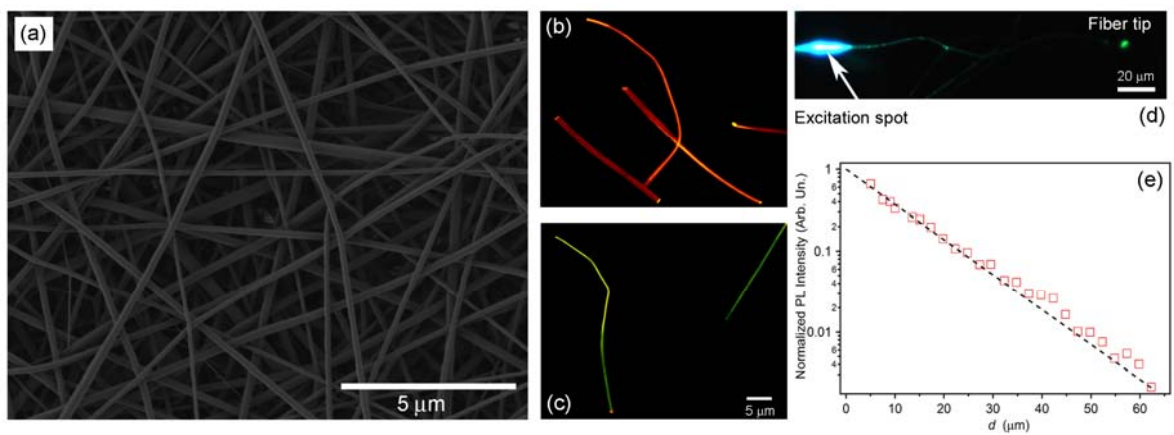

Figure 1. (a) SEM image of active electrospun polymer nanofibers in a randomly-oriented mat. (b)-(c) Fluorescence micrographs of isolated fibers embedding the laser dyes Rhodamine 6G (b) and Coumarin 334 (c), respectively. (d) Fluorescence micrograph of a light-emitting conjugated polymer fiber realized by electrospinning, highlighting the excitation spot and the bright and red-shifted emission from the tip. Light coupled to adjacent nanofibers, which are not excited directly, can be also appreciated in the micrograph. (e) Corresponding measured spatial decay of the light intensity (squares) guided along a single fiber, as function of the distance, $d$, of the observation point from the photoexcitation spot. The continuous line is the best fit of experimental data by an exponential function, Equation (1).

where *I* indicates the light intensity measured at a given point along the fiber, $I_0$ indicates the light intensity at the excitation point, and *d* is the distance between the two points along the fiber. For the most of investigated fibers, measured $\alpha$ values are between $10^2$ and $10^3$ cm$^{-1}$. For sake of comparison, we recall that these values are of the same orders of magnitude as those of the loss coefficients in semicrystalline conjugated polymer nanowires, realized by templating processes in the pores of anodic alumina membranes [14].

The functionality achievable by electrospun light-emitting nanofibers is not limited to waveguiding. Their flexibility and the possibility of greatly varying and controlling their chemical composition are an important added value. Depending on the used active material or blends, these nanostructures can be tailored to emit light in the whole visible [15] and in the near-infrared region [13], thus working as miniaturized light-emitting sources which can be potentially integrated in lab-on-a-chip devices [16]. Furthermore, we find that under pulsed optical pumping, increasing the excitation fluence leads to line narrowing of the emission, as typical of amplified spontaneous emission (ASE). ASE is due to the amplification of the radiation which is initially emitted spontaneously in the system and then travels along the active material. The needed population inversion is hence directly provided by external pumping, and ASE is favored by waveguiding of the radiation in the gain material, as often found in thin films which have a refractive index higher than the surrounding media. These effects are observed in electrospun light-emitting nanofibers as well, such as in PMMA fibers doped by Rhodamine. Differently from spincast films made of the same active materials, however, the ASE spectrum observed in these fibers is frequently structured, consisting in almost equally-spaced sharp peaks (full width at half maximum of 0.4 nm, mainly limited by our instrumental spectral resolution, as shown in the top-left inset of Figure 2a). The intensity of these peaks displays a well-behaved behavior upon increasing the excitation fluence, and a clear lasing threshold can be appreciated (at about 100 μJ cm$^{-2}$ in the characteristic input-output curve shown in Figure 2a). The lasing properties can be attributed to different mechanisms.

Part of the light emitted can be waveguided, back-reflected by the fiber tips and confined into individual fibers as in Fabry-Perot (*F-P*) microcavities. In fact, in some fibers we observe peaks whose spectral separation well correlates with the fiber length (*l*), taking into account the functional dependence of the mode spacing ($\Delta\lambda$) of a cylindrical *F-P* cavity:

$$\Delta\lambda \propto \lambda_0^2/2l, \qquad (2)$$

where $\lambda_0$ is the peak wavelength. Other optical modes can arise from surface whispery gallery modes in those electrospun fibers where this effect is enabled by the lack of structural defects. Finally, some electrospun fibers can display morphological defects along their axis, that can diffuse backward part of the light guided along the fibers, producing a set of optical cavities with random length. In these samples one can still observe a structured ASE emission, but the sharp peaks are not necessarily equally-spaced. Overall, such emission phenomena in electrospun nanofibers are indeed complex, depending not only on the fiber geometry but also on the orientational and supramolecular organization of macromolecules in the fibers, which can differ considerably from those in films due to high applied electric field and the stretching exerted in the electrospinning process. For these reasons, such properties are still subject of investigation aiming to assess the weights and photo-physics of the different possible contributions to the observed emission and to better exploit them for building low-cost lasing sources.

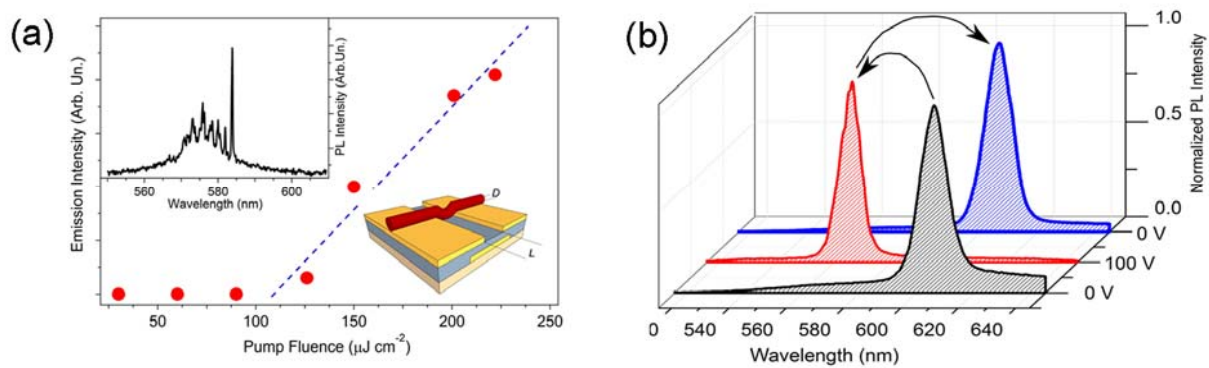

Figure 2. (a) Intensity of the sharp peak from lasing electrospun nanofibers vs pumping fluence. A threshold for lasing is clearly observed at an excitation fluence of about 100 μJ cm$^{-2}$. The dashed line is a guide for the eyes for above-threshold data. Upper-left inset: emission spectrum of an individual electrospun nanofiber. Pump fluence $\cong$ 200 μJ cm$^{-2}$. The bottom-right inset shows the scheme of a prototype field-effect transistor based on a bottom-contacted active polymer nanofiber utilizable for sensing. *D*: fiber diameter, *L*: active channel length. (b) Electrically-tunable ASE of a MEH-PPV thin film device embedding a nonlinear molecular layer. The sequence of ASE spectra is collected by sequentially applying 0 V, 100 V and 0 V, to interdigitated electrodes providing a local electric field of the order of $10^6$ V m$^{-1}$.

Many of these nanofibers are also conductive if realized with conjugation polymers [17-19]. Due to their combination of size downscaling, compositional flexibility, and controllable internal molecular ordering, such active fibers are also interesting both as tool to investigate fundamental conduction properties of semiconducting polymers and as active building-blocks to realize sensing elements based on transistor devices, as schematized in the bottom-right inset of Figure 2a. The very high surface available for sensing and the concomitant occurrence of light emission suggest that several different device architectures could be utilized in this framework, including organic light-emitting transistors [20, 21] and luministors [17, 22] based on the interplay of optical and electrical properties. In particular, the versatility of electrospinning in terms of fiber composition can enable the tunability of the emission by embedding further electro- or photo-active molecular systems in the fibers [23], as shown in Figure 2b, where we display the reversible ASE tunability, over 40 nm, of a conjugated polymer in a device based on nonlinear optical chromophores placed across interdigitated finger Cr/Au electrodes.

All these features (light-emission, waveguiding, lasing, etc.) could be exploited both at the level of individual fibers and at the scale of mats, whose active area can easily reach many $cm^2$ at laboratory scale. When dealing with flexible organic materials, the mechanical robustness of the overall medium is frequently matter of concern, especially for large-area samples. Fortunately, flexible electrospun mats show important advantages allowing one to overcome many issue related to mechanical fragility. With some polymers and a proper choice of the used solvents, one is able to obtain free-standing arrays in which nanofibers slightly overlap forming mutual joints which increase the mechanical robustness of the overall produced material. The formation of such joints is clearly ruled by the solvent evaporation kinetics during fiber deposition, and these peculiar structures can be directly imaged by SEM (Figure 3). We find that robust free-standing arrays of nanofibers, including lasing mats, can be obtained in this way, making these materials reliable for applications at macroscopic scales as recently demonstrated by the production of piezoelectric textiles which can be straightforwardly integrated in pressure and acceleration sensors [24]. More in general, nanofiber geometries are found to significantly improve mechanical properties compared to bulk materials, due to the disfavored formation of cracks, less frequent failures related to surface flaws, and, more importantly, due to their peculiar supramolecular organization [25, 26].

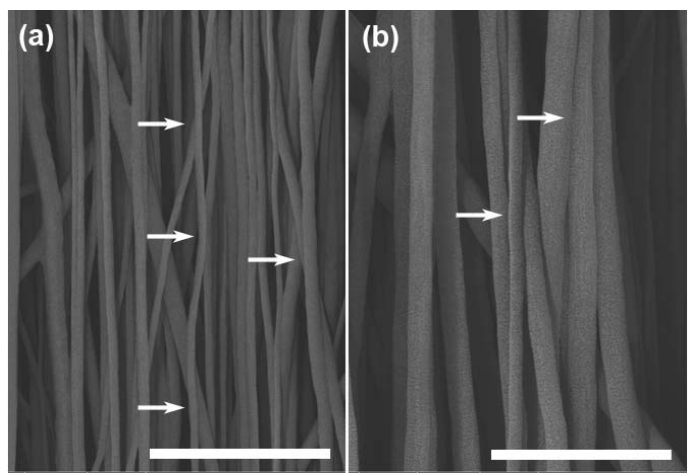

Figure 3. SEM micrographs at different magnifications of parallel, aligned electrospun fibers where mutual joints are formed during the deposition process. Arrays of aligned fibers are fabricated by using a rotating collector for deposition as described in Methods. The points of merging (joints) of nanostructures are highlighted by the horizontal arrows, and can involve two or more adjacent fibers. Scale bars, 5 μm (a) and 2 μm (b).

So far, these aspects have been rarely exploited for building mechanically stable photonic components based on active nanofibers. We anticipate that these properties can make electrospun polymer mats interesting candidates as light-emitting sources for building active textiles and smart surfaces with large area, for various technical applications.

## 4. CONCLUSIONS

In summary, electrospun polymer nanofibers can be used as light-emitting and lasing active media, with performances comparable or enhanced with respect to thin films. For instance, emission quantum yield is improved at least for some polymeric species [27]. Miniaturization, together with the remarkable tailorability of the structure of these fibers and consequently of their optical properties, can be other important advantages for the design and realization of new photonic components and, in perspective, for improved device integration. Importantly, polarized emission is also obtained from fibers with no need for additional rubbing or stretching. Another future development will be single mode emission, which could be achieved by patterning the fibers by imprinting lithographies [11], and emission properties can be further improved by realizing nanofibers under controlled atmosphere, thus decreasing the amount of quenching sites related to oxygen and humidity incorporated in the nanostructures during fabrication.

*Acknowledgments*. The research leading to these results has received funding from the European Research Council under the European Union's Seventh Framework Programme (FP/2007-2013)/ERC Grant Agreement n. 306357 (ERC Starting Grant "NANO-JETS"). V. Fasano is acknowledged for waveguiding measurements. Dr. F. Di Benedetto and Dr. P. Del Carro are also gratefully acknowledged for sample preparation.

## REFERENCES

[1] Grivas, C. and Pollnau, M., "Organic solid-state integrated amplifiers and lasers," Laser & Photon. Rev. 6, 419-462 (2012).
[2] Wiersma, D. S., "Disordered photonics," Nat. Photon. 7, 188–196 (2013).
[3] Yang, Y., Turnbull, G. A. and Samuel, I. D. W., "Hybrid optoelectronics: A polymer laser pumped by a nitride light-emitting diode," Appl. Phys. Lett. 92, 163306 (2008).
[4] Tsiminis, G., Ruseckas, A., Samuel, I. D. W. and Turnbull, G. A., "A two-photon pumped polyfluorene laser," Appl. Phys. Lett. 94, 253304 (2009).
[5] Quochi, F., "Random lasers based on organic epitaxial nanofibers," J. Opt. 12, 024003 (2010).
[6] O'Carroll, D., Lieberwirth, I., and Redmond, G., "Microcavity effects and optically pumped lasing in single conjugated polymer nanowires," Nature Nanotechnol. 2, 180 (2007).
[7] Reneker, D. H. and Chun, I., "Nanometre diameter fibres of polymer, produced by electrospinning," Nanotechnology 7, 216-223 (1996).
[8] Pisignano D., [Polymer Nanofibers], Royal Society of Chemistry, Cambridge (2013).
[9] Li D. and Xia, Y., "Electrospinning of nanofibers: Reinventing the wheel?," Adv. Mater. 16, 1151-1170 (2004).
[10] Camposeo, A., Persano, L. and Pisignano, D., "Light-emitting electrospun nanofibers for nanophotonics and optoelectronics," Macromol. Mater. Engineer. 5, 487-503 (2013).
[11] Di Benedetto, F., Camposeo, A., Pagliara, S., Mele, E., Persano, L., Stabile, R., Cingolani, R. and Pisignano, D., "Patterning of light-emitting conjugated polymer nanofibres," Nature Nanotech. 3, 614 (2008).
[12] Persano, L., Camposeo, A., Tekmen, C. and Pisignano, D., "Industrial upscaling of electrospinning and applications of polymer nanofibers: a review," Macromol. Mater. Engineer. 5, 504-520 (2013).
[13] Camposeo, A., Di Benedetto, F., Stabile, R., Cingolani, R. and Pisignano, D., "Electrospun dye-doped polymer nanofibers emitting in the near infrared," Appl. Phys. Lett. 90, 143115 (2007).
[14] O'Carroll, D., Lieberwirth I. and Redmond, G., "Melt processed polyfluorene nanowires as active waveguides," Small 3, 1178-1183 (2007).
[15] Camposeo, A., Di Benedetto, F., Cingolani, R. and Pisignano, D., "Full color control and white emission from conjugated polymer nanofibers," Appl. Phys. Lett. 94, 043109 (2009).
[16] Pagliara, S., Camposeo, A., Polini, A., Cingolani, R. and Pisignano, D., "Electrospun light-emitting nanofibers as excitation source in microfluidic devices," Lab Chip 9, 2851-2856 (2009).
[17] Tu, D., Pagliara, S., Camposeo, A., Persano, L., Cingolani, R. and Pisignano, D., "Single light-emitting polymer nanofiber field-effect transistors," Nanoscale 10, 2217-2222 (2010).

[18] Tu, D., Pagliara, S., Cingolani, R. and Pisignano, D., "An electrospun fiber phototransistor by the conjugated polymer poly[2-methoxy-5-(2 '-ethylhexyloxy)-1,4-phenylene-vinylene]," Appl. Phys. Lett. 98, 023307 (2011).
[19] Tu, D., Pagliara, S., Camposeo, A., Potente, G., Mele, E., Cingolani, R. and Pisignano, D., "Soft nanolithography by polymer fibers," Adv. Funct. Mater. 21, 1140-1145 (2011).
[20] Hepp, A., Heil, H., Weise, W., Ahles, M., Schmechel R. and von Seggern, H., "Light-emitting field-effect transistor based on a tetracene thin film," Phys. Rev. Lett. 91, 157406 (2003).
[21] Muccini, M., "A bright future for organic field-effect transistors," Nat. Mater. 5, 605-613 (2006).
[22] Dyreklev, P., Inganäs, O., Paloheimo J. and Stubb, H., "Photoluminescence quenching in a polymer thin-film field-effect transistor," J. Appl. Phys. 71, 2816-2820 (1992).
[23] Camposeo, A., Del Carro, P., Persano, L. and Pisignano, D., "Electrically tunable organic distributed feedback lasers embedding nonlinear optical molecules," Adv. Mater. 24, OP221-OP225 (2012).
[24] Persano, L., Dagdeviren, C., Su, Y., Zhang, Y., Girardo, S., Pisignano, D., Huang, Y. and Rogers, J. A., "High performance piezoelectric devices based on aligned arrays of nanofibers of poly[(vinylidenefluoride-co-trifluoroethylene]," Nat. Commun. 4, 1633 (2013).
[25] Arinstein, A., Burman, M., Gendelman O. and Zussman, E., "Effect of supramolecular structure on polymer nanofibre elasticity, "Nat. Nanotechnol. 2, 59–62 (2007).
[26] Pai, C.-L., Boyce M. C. and Rutledge, G. C., "Mechanical properties of individual electrospun PA 6(3)T fibers and their variation with fiber diameter," Polymer 52, 2295-2301 (2011).
[27] Morello, G., Polini, A., Girardo, S., Camposeo, A. and Pisignano, D., "Enhanced emission efficiency in electrospun polyfluorene copolymer fibers," Appl. Phys. Lett. 102, 211911 (2013).